# LOW IONOSPHERIC REACTIONS ON TROPICAL DEPRESSIONS PRIOR HURRICANES

Aleksandra Nina*[a], Milan Radovanović[b,c], Boško Milovanović[b], Andjelka Kovačević[d], Jovan Bajčetić[e], Luka Č. Popović[d,f]

[a]*Institute of Physics, University of Belgrade, Pregrevica 118, 11080 Belgrade, Serbia.*
[b]*Geographical Institute "Jovan Cvijić" Serbian Academy of Sciences and Art, Djure Jakšića 9, 11000 Belgrade, Serbia*
[c]*Perm National Research Polytechnic University, Perm, Russian Federation*
[d]*Faculty of Mathematics, University of Belgrade, Studentski trg 16, 11000 Belgrade, Serbia*
[e]*Department of Telecommunications and Information Science, Military Academy, University of Defence, Generala Pavla Jurišića Šturma 33, 11000 Belgrade*
[f]*Astronomical Observatory, Volgina 7, 11060 Belgrade, Serbia.*

**Abstract**

We study the reactions of the low ionosphere during tropical depressions (TDs) which have been detected before the hurricane appearances in the Atlantic Ocean. We explore 41 TD events using very low frequency (VLF) radio signals emitted by NAA transmitter located in the USA and recorded by VLF receiver located in Belgrade (Serbia). We found VLF signal deviations (caused ionospheric turbulence) in the case of 36 out of 41 TD events (88%). Additionally, we explore 27 TDs which have not been developed in hurricanes and found similar low ionospheric reactions. However, in the sample of 41 TDs which are followed by hurricanes the typical low ionosphere perturbations seem to be more frequent than other TDs.

*Corresponding author
  *Email addresses:* sandrast@ipb.ac.rs (Aleksandra Nina*),
m.radovanovic@gi.sanu.ac.rs (Milan Radovanović), b.milovanovic@gi.sanu.ac.rs
(Boško Milovanović), andjelka@matf.bg.ac.rs (Andjelka Kovačević),
bajce05@gmail.com (Jovan Bajčetić), lpopovic@aob.rs (Luka Č. Popović)

## 1. Introduction

The complexity of the physical and chemical processes in the atmosphere during the tropical depression and cyclone is caused by a number of processes, e.g. convective motion, thunderstorms, lightnings, strong electric fields between the clouds and the earth, air-sea interaction etc. These processes lead to numerous directly and indirectly caused changes at different altitude domains and geographical location. Some of them occur in the ionosphere and their analyses were in the focus of numerous previous scientific investigations. Thus, there are several studies of the tropical cyclone (TC) influence on ionosphere indicating electrical and electromagnetic effects (Isaev et al., 2002; Sorokin et al., 2005; Thomas et al., 2010). Also, the existence of acoustics and gravity waves (Xiao et al., 2007) indicates these feedback effects.

In cases of high ionosphere investigations, the analyses are based on calculations of the total electron content variations using Global Positioning System (GPS) technology (see for example Zakharov and Kunitsyn, 2012). Investigations based on the sudden disturbances in the low ionosphere and changes of the very low and low frequency (VLF/LF) radio signals which are used for its monitoring relate primarily to the short-term variations caused by lightning. There are two types of investigations: First, exploring the lightning activities intensification before a tropical depression (VLF signal variations are indicated as precursors in Price et al., 2007) and, second, analyzing the lightning activity during a hurricane (Peter and Inan, 2005; Thomas et al., 2010). Additionally, connections between the perturbations of the VLF/LF signals and meteorological factors are considered in several studies (Samsury and Orville, 1994; Molinari et al., 1994; Price et al., 2009).

However, after several decades of investigation, some of interactions between the tropospheric and ionospheric regions connected with TC have not been clarified and remain the subjects of numerous current multidisciplinary analyzes. In addition, there are no studies on some of these processes. One of them, considered in this study, relates to the low ionospheric perturbations in the period around tropical depression beginning (TDB) although they can be expected at lowest ionospheric altitudes because of significant changes in the low atmosphere.

Atmospheric research in this period can be of a great practical significance. Namely, it is known that the tropical depression could be developed into a tropical cyclone, but in some cases the troposphere returns to a unperturbed state. This difference leads to the important question: Whether there are any indications on the basis of which we could predict the time evolutions of atmospheric parameters after depressions? Certainly, one of the research directions that may provide some answers relates to the behavior of the low ionosphere, but relevant analyses have not been in the focus of previous investigations. The complexity of the atmospheric processes requires numerous analyses, but one of the first questions to be answered is whether there are changes in the low ionosphere in the periods of occurrence of depression preceding tropical cyclones. The study that we will present here relates just to this issue and to announces the pioneering research in this field.

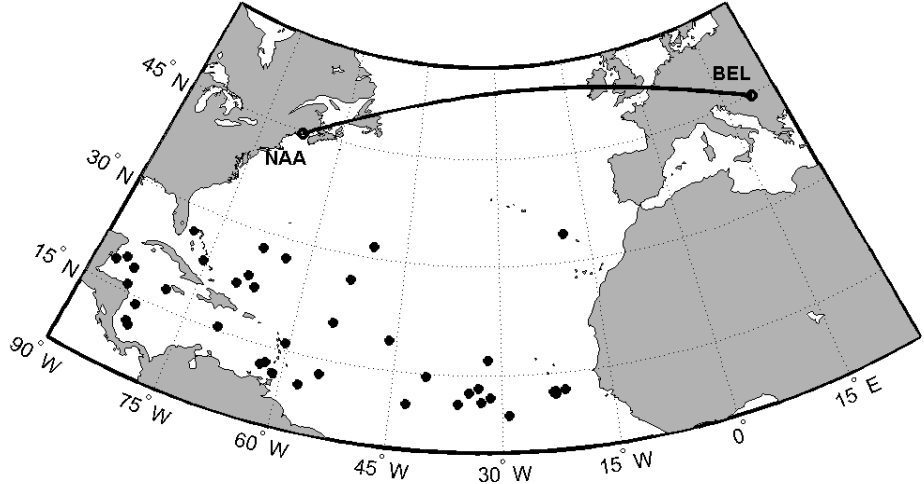

Figure 1: Locations of tropical depression beginnings (black circles) and great circle path of VLF signal emitted by NAA transmitter (USA) and detected by receiver located in Belgrade (Serbia).

The aim of this paper is to investigate possible long-term low ionospheric perturbations (i.e. perturbations lasting longer then those induced by lightning events which last up to a few minutes and which are in focus of numerous studies) in these periods. We investigated the effects of time of TDB and its geographical position with respect to considered low ionospheric part to detectability of the low ionospheric disturbances. In this study we used 24.0 kHz VLF radio signal emitted by NAA (USA) transmitter and registered by VLF receiver located in Belgrade (Serbia) to detect low ionospheric disturbances around 41 TDs before the registered hurricanes.

To find perturbations due to TD influence in the low ionosphere, first, we developed procedure for exceeding of any long-term signal amplitude deviations in the considered periods around a TDB, considering three referent days before the day of depression. After this statistical analysis, we take out typical amplitude shapes connected with these variations. In both cases, we considered subsamples regarding to recorded time of TDBs, and analysed sudden ionospheric disturbances (SIDs) with respect to time period and location of TDBs.

The paper is organized as follows: In Sec. 2 we describe our observations and experimental setup, in Sec. 3 we present a model for detection of SIDs. Results of our research concerning detections of the low ionospheric plasma perturbations using VLF signal analyses in periods of TDB events are presented in Sec. 4, and, finally, a short summary of this study is given in Sec. 5.

## 2. Observations and experimental setup

In this study, we base our analyses on data obtained in the low ionospheric observations by radio waves reflected from the considered medium. Because of characteristics of obtained information this experimental technique is used in a lot of studies related to the low ionosphere (including those mentioned in Introduction). Namely, numerous worldwide located transmitters and receivers allow detections of disturbances within relative large part of the low ionosphere and, consequently, it can be possible to detect unforeseen SIDs like those induced by solar flares or $\gamma$-ray bursts (Nina and Čadež, 2014; Šulić and Srećković, 2014; Nina et al., 2015) as well as located ones as, for example, those induced by solar terminator (ST) (Nina and Čadež, 2013). Also, continuously monitoring and receiving signals with time resolution of several tenth of milliseconds provide detection of very fast disturbances like those induced by lightnings (Inan et al., 1988).

Here, we used 24.0 kHz VLF radio signal emitted from NAA (USA) transmitter (44.65 N, 67.30 W) and registered by VLF/LF receiver located in Belgrade (Serbia) at geographical position (44.85 N, 20.38 E). As shown in Figure 1 this VLF signal propagates above the Atlantic Ocean at geographical latitude of about 44º (transmitter and receiver locations) to near 54º (latitude maximum of the considered signal great circle path (GCP)) and the recorded amplitude and phase variation can be caused by disturbances in these locations.

As show in Figure 1 and Table 1, the locations of the considered 41 TDs are at lower latitude than those of the GCP (from 9.7 N to 32.6 N). Consequently the presented analysis relates to possible connection of TDs with low ionospheric disturbances in the northern locations at middle latitude.

For this investigation we used a discrete set of data on amplitude variations of the signal registered by the AbsPAL (Absolute Phase and Amplitude Logger) receiver system with the sampling period of 1 min. The data related to tropical depressions are used from <http://weather.unisys.com/hurricane/> with time resolution of mostly 6 hours: at 3 h UT, 9 h UT, 15 h UT and 21 h UT. The analyzed events have occurred in the period 2004-2010 for which we have the recorded data by AbsPAL receiver.

## 3. Method and VLF signal analysis

As was noted above, the goals of this research are:

- Detection of SIDs in periods of TDB events;

- Their classification with respect to the amplitude deviation shape and occurrence time period;

- Examination of the influence of TDB time on considered detection and its characteristics, and

- Examination of the influence of TDB geographical location on considered detection and its characteristics.

We analyzed the sample of 41 TD events, and to extract disturbance in the period around TDBs we used the method based on the comparison of the signal amplitude detected during considered time period with its values within referent intervals in previous three days. Taking into account that TDBs occurred in the periods within 6 hours before the given times in Table 1 (time resolution of these data is 6 hours) we analyzed these periods, and periods of 6 hours before and after them (see at the case of TD Cindy occurred September 19, 2004 in Figure 2). In this way we investigate a possibility of ionospheric disturbances detection several hours before and after a TDB.

This time period choosing is applied to each individual event. However, diurnal variations in the atmosphere, occurrences of other sudden events and lack of data due to technical problems in monitoring required dividing of the whole sample in subsamples and extraction of some periods from consideration:

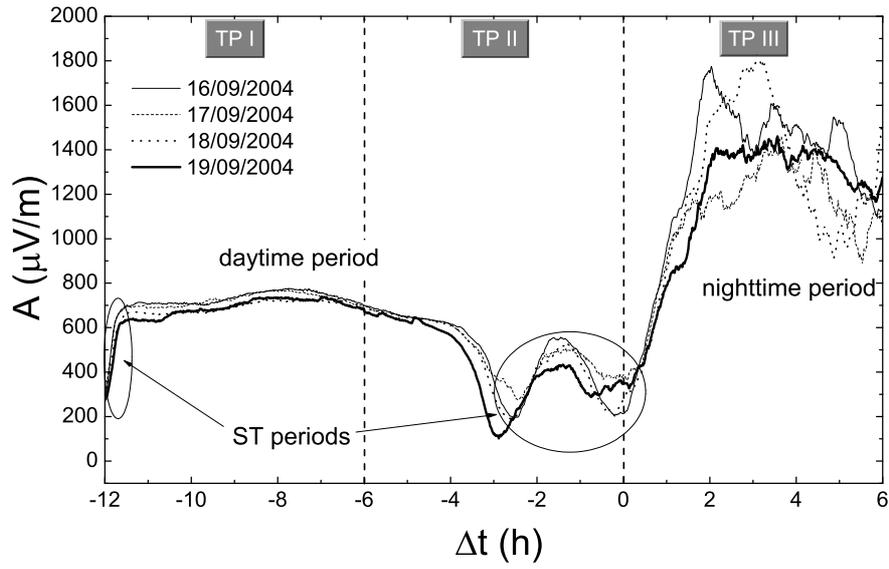

Figure 2: Signal amplitude time evolution $A(t)$ in time periods before TDB (TP I), within TDB (TP II) and after (TP III) TDB for the day of depression and three referent days before for the case of TD Lisa occurred on September 19, 2004. $\Delta t$ is time calculated from TDB given in Table 1. The marked periods represent time intervals when solar terminator (ST) affects medium within VLF signal propagate and show non-stationary properties of the low ionosphere between nighttime and daytime, and vice versa conditions.

**Subsamples.** The ionosphere properties relevant to propagation of electromagnetic waves are different during daytime and nighttime conditions (see Figure 2). Also, the propagation path of considered signal is very long and several hours during sunrise and sunset it is partially sunlit that results into a highly non-stationary behavior of the recorded VLF signal amplitude (Carpenter and Whitson, 1965; Clilverd et al., 1999). Therefore, to analyze the existing low ionosphere sudden disturbances in the period around TDBs, our sample is additionally divided into four subsamples regarding given TDBs at 3 h UT, 9 h UT, 15 h UT and 21 h UT. The number of events in these subsamples was 8, 2, 12 and 18, respectively.

**Relevant and non-relevant time periods.** During considered time periods, some other measurement setups detect occurrences of events which induced strong low ionospheric reactions. For example, GOES satellites detect several solar X-ray flares including very intensive one in the period around beginning of TD Katrina. Moreover, recently Vyklyuk et al. (2017) showed that there may be some connection between the solar activity and hurricanes. Solar activity as e.g. solar flares have a dominant role in the low ionospheric sudden disturbances and made that analysis of the influence of TDs on the ionosphere was impossible. For this reason, these time intervals are noticed as non-relevant period for the study and they are excluded from the analysis. Also, this reduction is applied for the cases when signal data are not good due to technical problems.

Procedure for detection of variation in signal characteristics and, consequently, in the low ionosphere, is dependent on SID intensity, duration and repetition. The recording of an event, like possible TDs, that do not intensively disturbs the ionosphere is a very complex task primarily because:

- Many phenomena can perturb the ionosphere (lightning, electron precipitation, solar activity, etc).

- The considered phenomenon (here, it is TD) and observed part of the ionosphere may have different characteristics during the similar event, especially considering the local perturbations.

- The shape of the signal related to the same perturbation can be different (Grubor et al., 2008).

- The shapes of amplitude variations connected with a low ionospheric perturbation depend on signal frequency, receiver location, daily period etc. It can be seen even in the cases of global and very strong SIDs like solar X-flares (Grubor et al., 2008).

In addition, there are no relevant studies about SIDs during periods of TDBs and we did not know what signal perturbations we could expect. For these reasons our study has two parts:

1. First, we made analyses in which we extracted any variations of signal amplitude from those expected from the relevant referent values.
2. Second, we extracted types of amplitude deviation that repeat in several cases in the first analysis.

*3.1. Method for detection of signal amplitude deviation*

This procedure is based on the comparisons of time evolutions of the standard deviation $\sigma$ in particular times for three referent days before the day of depression, $\sigma_1(t)$, and for all four days $\sigma_2(t)$ which are given by expressions:

$$\sigma_N(t) = \sqrt{\frac{1}{n_N(t)} \sum_{i=1}^{n_N(t)} (A_i(t) - A_{mN}(t))^2}, \quad (1)$$

where $N = 1$ refers to the three day, and $N = 2$ to the four day period. $A_{mN}(t)$ are mean values of considered amplitudes at time $t$ while $n_N(t)$ are the numbers of amplitude values included in calculations of their mean values at time $t$. The deviation of signal in the day of depression at time $t$ is quantified by coefficient $r(t)$ defined by relation:

$$r(t) = \frac{\text{abs}(\sigma_2(t) - \sigma_1(t))}{\sigma_1(t)}. \quad (2)$$

To record the time when the signal deviation i.e. SID is important for the analysis we introduce the following criteria:

- It is necessary that amplitude for the day of a depression is relevant for analysis and, consequently, included in calculations for the considered time $t$.

- The values for $n_1(t)$ and $n_2(t)$ have to be at least 2 and 3, respectively. Thus we allow possibility that one of reference signals in the considered time $t$ is non-relevant.

- The deviation of signal at time $t$ in the day of a depression is significant if $r(t) \geq 100\%$.

- Deviation for the TD event is recorded if at least 50% of values $r(t)$ is significant within one hour.

The examples of the events for which applying of these criteria results in the absence and existence of SID detections are visualized in Figure 3 for hurricanes Paloma (left panel) and Lisa (right panel), respectively. Here signal amplitudes for the considered days are presented in the upper panels, while $r$ calculated using Eq. 2 is given in the bottom panels where solid thick lines indicate times when the given conditions are satisfied. As we can see they exist only in the case of hurricane Lisa. The comparison of the upper and bottom right panels shows that marked times start before and finish after visible signal deviation which is expected because of the last criterium. However, the second example indicates important influence of $\sigma_2$ value. As it can be seen in this case its small values result in "mathematical" detection of signal variation which is not real important (period before the marked intervals with "real" considered signal deviations). For this reason we have additionally done visual check of detected SIDs and, because of very small $\sigma_2$ or very large $\sigma_1$ in some extracted cases, we excluded or included several cases, respectively. Also we included one case of the evident reaction that was not recorded by program (it did not satisfy the last criterion) because it began toward the end of considered interval and its large part that occurred later was not processed.

To validate and classify the intensity of extracted deviations using previous procedure we calculated the deviations of amplitude values in these periods $A_D$ around tropical depression beginnings with respect to the relevant mean values calculated for referent, quiet, days $A_{m1}$. Because of the more adequate comparisons of these deviations in different time periods we divided the obtained values with standard deviations for referent days $\sigma_1$ and expressed them as reduced variations $\delta_D$:

$$\delta_D(t) = \frac{abs(A_D(t) - A_{m1}(t))}{\sigma_1(t)}. \tag{3}$$

This reduction allow us to compare the deviations during, for example, the nighttime and daytime periods when the variations usually different in the recorded signal amplitude are different.

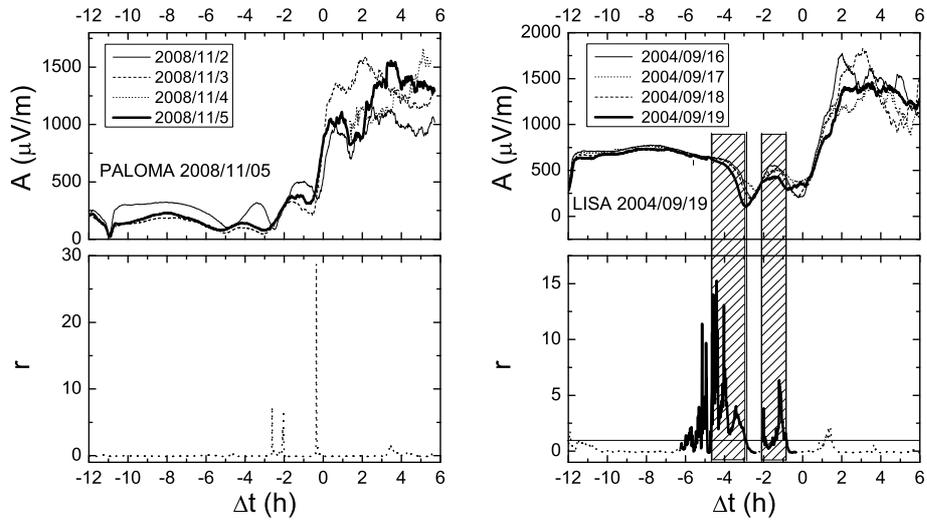

Figure 3: Examples of the events without and with SID detection. The signals for four days related to Paloma and Lisa hurricanes are shown in the upper left and right panels respectively. The obtained values $r$ using Eq. 2 are shown in the bottom panel. Times when noticed criteria for SID detection are satisfied can be visualized by solid tick lines existing only in the second case while intervals of "real" SID detections are marked by shaded areas. Horizontal lines indicate values $r = 1$.

*3.2. Extraction of signal deviation types*

Regarding the extracted periods by described method and visual analyses, we noticed three types of amplitude variation which are shown in Figure 4.

**Type 1.** This type of amplitude changes is characterized by larger amplitude decreasing in daytime periods respect to expected time evolution shape obtained from referenced amplitude values (upper panel). In most cases, these decreases started just before sunset periods and the same tendencies were continued in period of partially sunlit signal propagation path. In other words, characteristic signal amplitude shapes in sunset periods started earlier in the TD day.

**Type 2.** Its characteristic is amplitude decrease or saturation instead of expected convex shape (middle panel) when a ST affect the VLF signal propagation path (usually during sunset).

**Type 3.** In this case, amplitude starts earlier to increase from the dip in ST periods to nighttime values. So, as shown in the bottom panel of Figure 4, larger amplitude values than those expected are recorded in this period.

These three types are analyzed in the second part of the study (see Subsection 4.2).

## 4. Results and Discussions

In order to have clear presentation of obtained results we divided this section in two parts: the first part of this study is based on signal variation detections in the periods around TDBs using criteria described in the Section 3 while determination of a typical shape of amplitude variations is presented in the second part of the study. In both cases we visualized detected signal variations at TDB geographical locations and show results for four subsamples which were formed according to TDBs. As was noted above, the latter classification is very important because of very different atmospheric properties during daytime, nighttime and ST periods.

In following text we show results of these studies and discuss them.

*4.1. Detections of signal amplitude deviations*

To extract amplitude variations we applied the method described in the Section 3 on each individual case in the considered sample of 41 TD events. First, we studied whole time period and indicated the detected variations in Figure 5. After this, we focused our analysis on periods before, during and after TDBs. For this reason, we divided time interval on three parts and show results for each of them separately in Table 2.

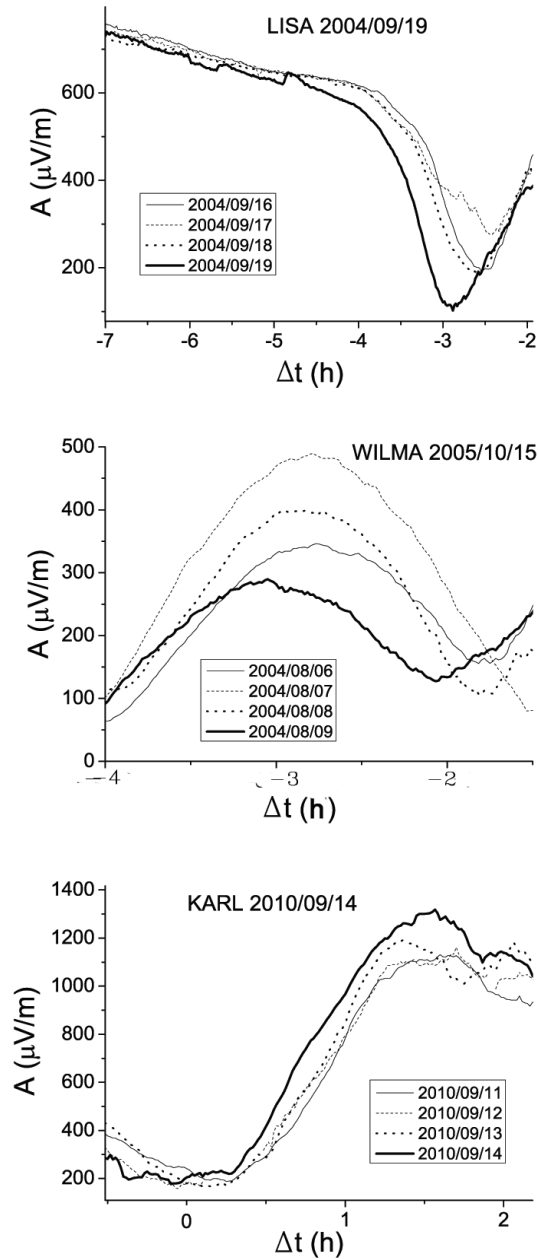

Figure 4: Three types of amplitude changes several hours around depression beginnings. Upper panel: amplitude decrease during daytime conditions (usually immediately before ST period), middle panel: amplitude decrease during ST period before nighttime (in this case there were two TDB events), and bottom panel: amplitude increase at the end of ST period before nighttime conditions. The time 0 h refers to the time of TDBs.

*4.1.1. Whole period*

The presence of signal deviations recorded within all considered periods of 18 h is shown in Figure 5 where panels relate to subsamples formed with respect to TDBs. Here, the nighttime perturbations are indicated by black squares, open circles relate to the daytime perturbations while × represent perturbations during ST periods. The cases without recorded deviations are marked as small points. From our signal exploration we can outline several results:

- We detected ionospheric disturbances in 36 of 41 TD events (88%).

- The latitudes of TDBs for all 5 events for which low ionospheric responses are not recorded are below latitude of 15 degrees.

- Signal deviations are present in a significant number of TD events in all cases beside the TDs registered at 9 UT (in this subsample we have only two cases that do not allow us to give any conclusion): 100% for TD beginnings at 3 h UT, 92% for TD beginnings at 15 h UT, and 83% for TD beginnings at 21 h UT.

- Signal amplitude deviations are detected during all three time periods of day (daytime, nighttime and ST periods).

- In several TD events we detected considered perturbations in two or all three periods, which indicates that measurable disturbances can be repeated.

Note here that some relevant ionospheric feedbacks are not detected because they occurred in non-relevant period (see Section 3) or because of shorter period of signal deviation than one required by the given relevant criterion. So, the number of presented detection is the minimum number of SIDs in the considered periods.

To validate and classify the intensity of extracted deviations using previous procedure we calculated the reduced variations $\delta_\mathrm{D}$ given by Eq. (3). The mean of values for extracted periods refer to particular TD events and they are shown in Figure 5. All of them are larger than 3 except for TDs Vince (2.7) and Willma (2.5) which are noticed as the events with detected deviation because of clearly visible variations (they are not detected by used procedure due to large $\sigma_1$). This result indicates large significance of extracted deviations. In Figure 5 we grouped the obtained reduced variations in ranges of $\delta_D$ below 5, between 5 and 10, and above 10. In considered subsamples one can see that the deviations are usually intensive for the events with TDBs at 3 UT and 15 UT, while their $\delta_D$ values are in all three ranges for the events with TDBs at 21 UT.

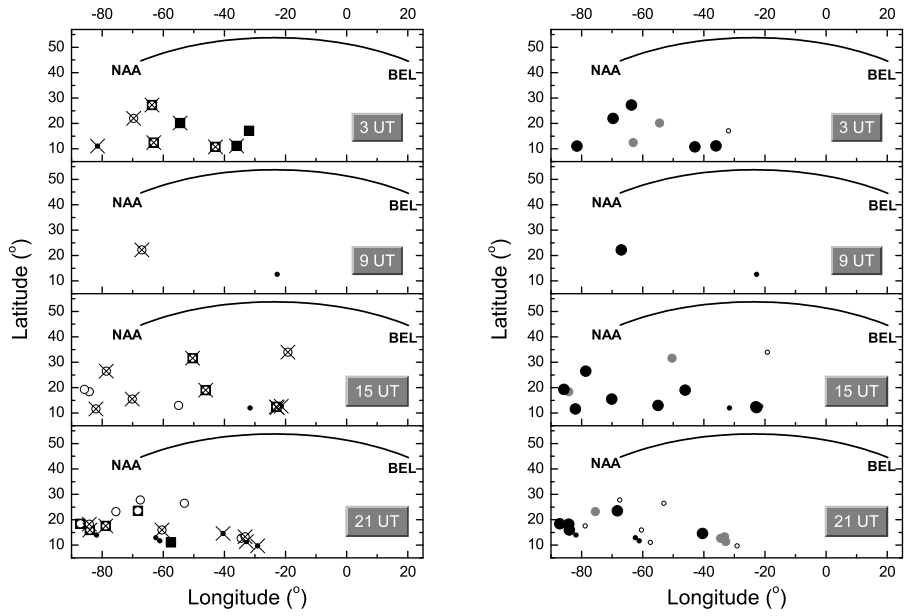

Figure 5: Locations of depression beginnings. Panels from the top to the bottom relate to depressions beginning at 3 UT, 9 UT, 15 UT and 21 UT. Left panels: Detection of VLF signal perturbation during all considered time period for relevant depression is indicated by black squares (nighttime perturbations), open circles (daytime perturbations) and x (perturbations during ST periods). TDs without recorded low ionosphere reactions are shown as dots. Right panels: reduced variations $\delta_D$. Small open circles relate to $\delta_D < 5$, gray circles are for $5 < \delta_D < 10$ and large full circles show $\delta_D > 10$.

*4.1.2. Periods before, during and after TDBs*

One of the goals of this study is to examine the possibility of prediction of TDBs using ionospheric perturbations and to analyze SIDs after TDBs. To do that we divided the analysis in three 6 hour time periods (TPs) (see Figure 2): TP I before TDBs, TP II within TD starts, and TP III after TDBs, and applied the procedure given in the Section 3 on four subsamples as in the previous analysis. From results of this analysis represented in Table 2 we can outline:

- There is evidence for perturbation of the low ionosphere for all three relevant subsamples for all three periods. Statistically, the most important are variations before TDs (in TP I) when recorded beginnings were at 15 h UT (77%), in TP II for TDBs at 3 h UT (88%) and in TP III for TDBs at 15 h UT (84%).

- Dominant nighttime reactions are recorded for the TDBs in 3 h UT for TP II (75%).

- Significant number of reactions during the daytime period is recorded for the TDBs at 15 h UT and 21 h UT.

- Reactions in the ST periods are not recorded only for the TDBs at 9 h UT.

*4.2. Types of amplitude variations*

In pervious analysis we considered all variations that satisfy required conditions given in the Section 3. Now, in the second part of this study, we investigate the existence of typical signal shapes that can be connected with TDs. Also, we analyze the influence of TDB locations and signal deviation time periods to extract particular shapes.

As mentioned in the Section 3 the visual analysis of the reported events indicates that we can mark off three types of amplitude variations shown in Figure 4. To study occurrence periods for these three types of reaction, we considered all visible relevant variations and again analyzed subsamples. The obtained results, given in Table 3, show that at least one of indicated SID types is visible in 6 of 8 (75%), 1 of 2 (no good statistics), 9 of 13 (69%; here analysis of Type 3 is possible only in 4 cases - see comment below) and 17 of 18 (94%) TD events for subsamples related to TDBs at 3 UT, 9 UT, 15 UT and 21 UT, respectively. As it can be seen from presented data, the most represented amplitude changes relate to Type 1 (recorded in 25 of 41 cases e.g. 61%) while variations of Type 2 and Type 3 are visible in 15 of 41 (37%) and 11 of 32 (34%) cases, respectively. In the last case, total number of the considered TD events is 32 because analysis of Type 3 is possible only in 4 cases in subsamples of TD beginnings at 15 h UT.

The SIDs of Type 1 is noted before TDBs, in all three TPs. Type 2 is usually recorded in TP II at TDBs in 21 h UT. Finally, Type 3 is visible in most cases in TP II for TDBs at 3 h UT and TP III for TDBs at 21 h UT. It is important to point out that the focus of our study is on the determination of changes in the periods around TDBs. So, one of the open questions in this study is examination of existence of these types in longer time period that will be the focus of our further investigations. Here this is primarily important for the cases of Type 3 for TDBs at 15 h UT because the considered time period does not include intervals when Type 3 appears in 9 of 13 cases.

From the locations of TDs which can be connected with some of the considered signal deviation types, marked as × in Figure 6, we can conclude that it is not noticeable their spatial grouping. This result indicates that the large geographic areas are mutually coupled at the tropospheric and ionospheric heights in periods around TDBs.

The presented analysis shows that the considered types of changes in the VLF signal occur in most of the observed cases which are related to depression prior hurricanes. However, the processes that develop from depression to hurricane are very complicated, and hurricanes do not necessarily occur after a TD occurrence. Consequently, a very important question is: whether the deviations extracted in this study can be an indicator which shows whether the depression passes into a hurricane or if the weather stabilizes? For this reason, we additionally analyzed the sample of 27 TDs that did not develop into a hurricane (see Table 4). The detected deviations (all three types are recorded) are shown in Figure 7 where one can see that the low ionospheric responses was noted in 15 cases (56%). We particularly detected the considered variation types in 4 of 7 (57%), 0 of 1 (no good statistics), 5 of 8 (62%) and 6 of 11 (55%) events for subsamples related to the TDBs at 3 UT, 9 UT, 15 UT and 21 UT, respectively. The obtained values show that we cannot a priori predict the formation of a hurricane from the analysis of the noticed three types of the low ionospheric perturbations. However, the total percentage is smaller in comparison with the first one obtained for the sample of depressions prior hurricanes (80%). This indicates the need to further detailed analyzes of this difference which will be in the focus of our future investigation.

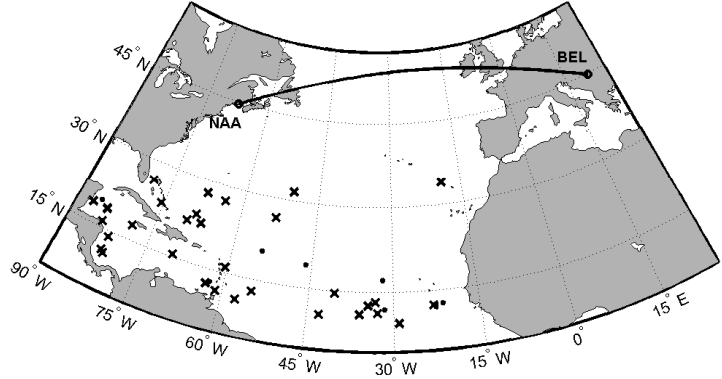

Figure 6: The same as in Figure 1 but with marked TD events connected with at least one Type of SID as ×. Locations of TDs without detected relevant signal variations are marked as •.

Explanations of relationships between the intensive stratospheric events and the ionospheric perturbations are not trivial because of very complex processes and their causal links in the atmosphere. Investigations show that the VLF/LF signals as "probe" for the low ionospheric monitoring are sensitive to variations of atmospheric pressure, humidity, wind velocity and temperature (Rozhnoi et al., 2014). However, even in the case of tropical cyclones as one of the strongest stratospheric instabilities, investigations of their connections with the ionospheric sudden disturbances are the subject of contemporary research. The studies primarily indicate two types of processes which can be considered as possible mechanisms of the analyzed connection: atmospheric waves and electric field disturbance arises due to perturbation in the atmosphere-ionosphere electric circuit. In the case of the first type, a broad spectrum of internal gravity waves (IGWs), tides and planetary waves are analyzed as the source of the ionospheric disturbances (Kazimirovsky, 2002; Dhaka et al., 2003; Sharkov, 2012). On the other side, there is a theory which explains the upper atmosphere variations by the penetration into the ionosphere of the electric field from the electric charge formed in the stratosphere during a TC (Pulinets et al., 2000; Sorokin et al., 2005). Although the changes are larger in the atmospheric parameters (pressure, temperature, wind speed) during TCs than in periods of TDs their properties are similar. For this reason, the possible role of the atmospheric waves and electric activity in troposphere-ionosphere coupling in the periods of TDBs requires a detailed research and represents the open questions for the upcoming investigations.

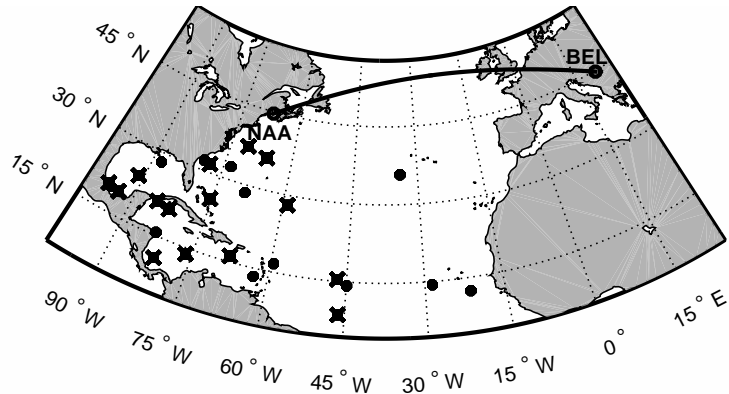

Figure 7: The same as in Figure 6 but for the TDs which have not been developed in hurricanes.

## 5. Conclusions

This is the first work that investigates sudden disturbances in the low ionosphere in the periods around TDBs. The most important result of this study is that we found responses of the low ionosphere that lasts at least several tens of minutes in the periods of several hours around TDBs for near 90% of the considered cases which indicate large possibility of connection between the troposphere and low ionosphere in the periods around TDBs.

The ionospheric disturbances recorded by 24 kHz VLF signal emitted by NAA transmitter in the USA and received by AbsPAL receiver in Serbia are detected at geographic latitude between about 44º and 54º which relates to locations northern than those where considered 41 TDs occurred.

From the presented analysis we can outline the following conclusions:

1. The low ionospheric perturbations are recorded in 88% cases. The latitudes of TDBs for which low ionospheric responses are not recorded are below latitude of 15 degrees.
2. Signal deviations are present in significant number of TD events in all cases of TDBs beside the TDs registered at 9 UT (in this subsample we cannot give any conclusion because of bad statistics).
3. Signal amplitude deviations are detected during all three time periods of day (daytime, nighttime and ST periods) and they can be repeated.
4. Signal amplitude deviations are detected at different times in relation to TDBs indicating a possibility of detecting SIDs as precursor of TDs.
5. We noticed three types of signal amplitude changes: its decreasing during day (usually with beginning before sunset and ending during sunset), decreasing in ST periods and increasing before nighttime conditions. Variations in nighttime atmosphere have not preferred shape that we can extract from considered sample.
6. Similar as in the case of TDs prior hurricanes, the three characteristic types of signal amplitude changes are also recorded in the case of TDs which have not been developed in hurricanes.

This analysis expands previous relevant studies in two aspects:

- The presented work relates to the VLF signal which propagates northern than considered TDs preceding to hurricanes. In this way, it extends the investigations of the VLF radio signal perturbation before hurricanes given in Price et al. (2007) relating to atmosphere above Africa.

- Keeping in mind that relevant low ionospheric investigations usually consider short-term effects of lightnings (amplitude peaks below 1 s lasting, and several minutes duration of perturbations induced by lightning induced electron precipitations), considerations of amplitude changes of several tenths of minutes are expanding of considered low ionosphere disturbance lasting.

Comparing the sample of 41 TDs followed by hurricanes with the sample of 27 TDs which have not been followed by these events we found that a slightly higher frequency of the signal perturbation in the first sample than in the second one. This may indicate some specific perturbations in the low ionosphere prior hurricanes that should be investigated. Consequently, a detailed analysis of TD influence on the ionosphere is needed. This study will be in the focus of our upcoming research, primarily with the aim to analyze the possibility of detecting differences in the low ionosphere that could be related to the further evolution of the tropical depression in hurricanes or restoration the troposphere in a calm state. In addition, we want to point out that presented procedure for extraction of signal amplitude deviation of expected values obtained from referent days can be applied to different events.

## Acknowledgments

The data for this paper related to tropical depressions is available at Unisys Weather Information Systems http://weather.unisys.com/hurricane/. Requests for the VLF data used for analysis can be directed to the corresponding author.

The authors are thankful to the Ministry of Education, Science and Technological Development of the Republic of Serbia for the support of this work within the projects III 44002, 176001, 176002, III 47007 and TR 32030. Also, this study is made within the VarSITY project and COST project TD1403.

Table 1: Geographical (latitude and longitude) and temporal (date and hour) characteristics of TD beginnings. Data were taken from the site http://weather.unisys.com/hurricane/.

| No | No (for subsample) | Latitude (°) | Longitude (°) | Date (year/month/day) | Hour (UT) | Hurricane Name |
|----|----|----|----|----|----|----|
| | | | | **3 UT** | | |
| 1 | 1 | 27.3 | -63.7 | 2010/10/29 | 3 | Shary |
| 2 | 2 | 22 | -69.7 | 2005/9/18 | 3 | Rita |
| 3 | 3 | 20.2 | -54.5 | 2006/9/11 | 3 | Gordon |
| 4 | 4 | 17.1 | -31.9 | 2010/9/21 | 3 | Lisa |
| 5 | 5 | 12.5 | -63.1 | 2005/7/5 | 3 | Dennis |
| 6 | 6 | 11.2 | -36 | 2004/8/25 | 3 | Frances |
| 7 | 7 | 11.1 | -81.5 | 2005/10/27 | 3 | Beta |
| 8 | 8 | 10.8 | -42.9 | 2005/7/11 | 3 | Emily |
| | | | | **9 UT** | | |
| 9 | 1 | 22.2 | -67 | 2010/10/6 | 9 | Otto |
| 10 | 2 | 12.6 | -22.7 | 2008/7/3 | 9 | Bertha |
| | | | | **15 UT** | | |
| 11 | 1 | 34 | -19.2 | 2005/10/9 | 15 | Vince |
| 12 | 2 | 31.6 | -50.4 | 2005/11/29 | 15 | Epsilon |
| 13 | 3 | 26.5 | -78.6 | 2005/9/6 | 15 | Ophelia |
| 14 | 4 | 19.3 | -85.8 | 2005/10/1 | 15 | Stan |
| 15 | 5 | 19 | -46.1 | 2005/9/1 | 15 | Maria |
| 16 | 6 | 18.4 | -84.2 | 2008/7/20 | 15 | Dolly |
| 17 | 7 | 15.5 | -70.1 | 2008/8/25 | 15 | Gustav |
| 18 | 8 | 13 | -55 | 2005/9/17 | 15 | Philippe |
| 19 | 9 | 12.7 | -21.4 | 2010/9/12 | 15 | Julia |
| 20 | 10 | 12.5 | -23 | 2006/9/12 | 15 | Helene |
| 21 | 11 | 12.2 | -22.7 | 2004/8/13 | 15 | Danielle |
| 22 | 12 | 12 | -31.6 | 2007/8/13 | 15 | Dean |
| 23 | 13 | 11.6 | -82 | 2009/11/4 | 15 | Ida |
| | | | | **21 UT** | | |
| 24 | 1 | 27.8 | -67.5 | 2005/9/5 | 21 | Nate |
| 25 | 2 | 26.5 | -53.1 | 2006/9/27 | 21 | Isaac |

| 26 | 3  | 23.5 | -68.3 | 2008/9/25  | 21 | Kyle     |
|----|----|------|-------|------------|----|----------|
| 27 | 4  | 23.2 | -75.5 | 2005/8/23  | 21 | Katrina  |
| 28 | 5  | 18.4 | -87.1 | 2005/7/3   | 21 | Cindy    |
| 29 | 6  | 18.3 | -84.2 | 2010/9/14  | 21 | Karl     |
| 30 | 7  | 17.6 | -78.8 | 2005/10/15 | 21 | Wilma    |
| 31 | 8  | 16   | -84   | 2010/10/11 | 21 | Paula    |
| 32 | 9  | 16   | -60.4 | 2004/9/13  | 21 | Jeanne   |
| 33 | 10 | 14.6 | -40.4 | 2006/9/3   | 21 | Florence |
| 34 | 11 | 14   | -81.8 | 2008/11/5  | 21 | Paloma   |
| 35 | 12 | 13.3 | -33.2 | 2004/9/19  | 21 | Lisa     |
| 36 | 13 | 12.9 | -62.4 | 2006/8/24  | 21 | Ernesto  |
| 37 | 14 | 12.7 | -34.5 | 2005/8/4   | 21 | Irene    |
| 38 | 15 | 11.7 | -61.1 | 2004/8/9   | 17 | Charley  |
| 39 | 16 | 11.4 | -32.8 | 2004/9/6   | 21 | Karl     |
| 40 | 17 | 11.1 | -57.5 | 2010/10/29 | 21 | Tomas    |
| 41 | 18 | 9.7  | -29.1 | 2004/9/2   | 21 | Ivan     |

Table 2: Number of detected SIDs in daytime, nighttime and ST periods during considered TPs and subsamples. The SIDs beginning during daytime and ending in the sunset are considered as daytime events. The results for all TPs are shown at the bottom for subsamples and in total.

| TDB | TD | nighttime | daytime | ST | TD with SID | |
|---|---|---|---|---|---|---|
| TP I | | | | | | |
| 3 h  | 8  | 0 | 3 | 2 | 4  | 22 (54 %) |
| 9 h  | 2  | 0 | 0 | 0 | 0  | |
| 15 h | 13 | 1 | 2 | 8 | 10 | |
| 21 h | 18 | 0 | 8 | 0 | 8  | |
| TP II | | | | | | |
| 3 h  | 8  | 6 | 0 | 3 | 7  | 25 (61%) |
| 9 h  | 2  | 0 | 0 | 0 | 0  | |
| 15 h | 13 | 0 | 7 | 1 | 7  | |
| 21 h | 18 | 0 | 7 | 9 | 11 | |
| TP III | | | | | | |
| 3 h  | 8  | 0 | 0 | 5 | 5  | 27 (66%) |
| 9 h  | 2  | 0 | 1 | 0 | 1  | |
| 15 h | 13 | 1 | 7 | 7 | 11 | |
| 21 h | 18 | 5 | 0 | 6 | 10 | |

| TDB | 3 h | 9 h | 15 h | 21 h | Total |
|---|---|---|---|---|---|
| All TPs | 8 (100%) | 1 (50%) | 12 (92%) | 15 (83%) | **36 (88%)** |

Table 3: Number of detected SIDs of Types I, II and III during considered TPs and subsamples. The results for all types are shown at the bottom for subsamples and in total.

| TDB | TD | TP I | TP II | TP III | TD with SID | |
|---|---|---|---|---|---|---|
| Type 1 | | | | | | |
| 3 h | 8 | 5 | 0 | 0 | 5 | |
| 9 h | 2 | 0 | 0 | 1 | 1 | 25 (61%) |
| 15 h | 13 | 0 | 2 | 6 | 8 | |
| 21 h | 18 | 6 | 7 | 0 | 11 | |
| Type 2 | | | | | | |
| 3 h | 8 | 0 | 1 | 2 | 3 | |
| 9 h | 2 | 0 | 0 | 0 | 0 | 15 (37%) |
| 15 h | 13 | 0 | 0 | 4 | 4 | |
| 21 h | 18 | 0 | 7 | 1 | 8 | |
| Type 3 | | | | | | |
| 3 h | 8 | 1 | 3 | 0 | 4 | |
| 9 h | 2 | 0 | 0 | 0 | 0 | 11 (34%) |
| 15 h | 4 | 0 | 0 | 0 | 0 | |
| 21 h | 18 | 0 | 1 | 6 | 7 | |

| TDB | 3 h | 9 h | 15 h | 21 h | **Total** |
|---|---|---|---|---|---|
| All Types | 6 (75%) | 1 (50%) | 9 (69%) | 17 (94%) | **33 (80%)** |

Table 4: The same as in Table 1 but for TD which have not developed in hurricane. Data were taken from the site http://weather.unisys.com/hurricane/.

| No | No (for subsample) | Latitude (°) | Longitude (°) | Date (year/month/day) | Hour (UT) | Hurricane Name |
|---|---|---|---|---|---|---|
| **3 UT** | | | | | | |
| 1 | 1 | 11.1 | -81.5 | 2005/10/27 | 3 | Beta |
| 2 | 2 | 13.5 | -62.7 | 2005/11/14 | 3 | Twenty_seven |
| 3 | 3 | 16.6 | -59.4 | 2006/8/1 | 3 | Chris |
| 4 | 4 | 36.1 | -66 | 2007/7/31 | 3 | Chantal |
| 5 | 5 | 23.9 | -91.1 | 2007/8/15 | 3 | Erin |
| 6 | 6 | 31.9 | -79.6 | 2008/7/19 | 3 | Cristobal |
| 7 | 7 | 19.9 | -95.7 | 2005/6/28 | 22 | Bret |
| **9 UT** | | | | | | |
| 8 | 1 | 35.8 | -34.1 | 2004/9/9 | 9 | Ten |
| **15 UT** | | | | | | |
| 9 | 1 | 28.1 | -59 | 2005/10/8 | 15 | Twenty_two |
| 10 | 2 | 16.1 | -68 | 2005/10/22 | 15 | Alpha |
| 11 | 3 | 21.1 | -85.3 | 2006/6/10 | 13 | Alberto |
| 12 | 4 | 32.5 | -73.4 | 2006/7/18 | 15 | Beryl |
| 13 | 5 | 15.6 | -83 | 2008/10/14 | 15 | Sixteen |
| 14 | 6 | 37.3 | -71 | 2009/5/28 | 15 | One |
| 15 | 7 | 14.4 | -28.6 | 2009/8/11 | 10 | Ana |
| 16 | 8 | 20.6 | -82.5 | 2010/9/28 | 15 | Nicole |
| **21 UT** | | | | | | |
| 17 | 1 | 8.9 | -46.2 | 2004/8/13 | 21 | Earl |
| 18 | 2 | 31.6 | -78.1 | 2004/8/27 | 21 | Gaston |
| 19 | 3 | 25.3 | -75.4 | 2005/7/21 | 21 | Franklin |
| 20 | 4 | 19.4 | -93.2 | 2005/7/23 | 21 | Gert |
| 21 | 5 | 28.5 | -68.7 | 2005/8/2 | 21 | Harvey |
| 22 | 6 | 14.3 | -44.9 | 2005/8/13 | 21 | Ten |
| 23 | 7 | 19.5 | -95 | 2005/8/22 | 16 | Jose |
| 24 | 8 | 15.4 | -46.8 | 2005/8/28 | 21 | Lee |
| 25 | 9 | 12.5 | -21.5 | 2006/8/21 | 21 | Debby |
| 26 | 10 | 28.2 | -88.1 | 2008/8/3 | 21 | Edouard |

| 27 | 11 | 13.9 | -76.2 | 2010/9/23 | 18 | Matthew |